\documentclass{article}
\usepackage[T1]{fontenc}
\usepackage[utf8]{inputenc}
\usepackage{ismir} % Remove the "submission" option for camera-ready version
\usepackage{amsmath,cite,url}
\usepackage{graphicx}
\usepackage{xcolor}

\usepackage{multirow}

\title{Barwise Section Boundary Detection in Symbolic Music Using Convolutional Neural Networks}

\twoauthors
   {Omar Eldeeb} {Technical University of Munich \\ \texttt{omar.eldeeb@tum.de}}
   {Martin E.\ Malandro} {Sam Houston State University \\ \texttt{malandro@shsu.edu}}

\def\authorname{O. Eldeeb and M. E. Malandro}

\begin{document}

\maketitle

\begin{abstract}
Current methods for Music Structure Analysis (MSA) focus primarily on audio data. While symbolic music can be synthesized into audio and analyzed using existing MSA techniques, such an approach does not exploit symbolic music's rich explicit representation of pitch, timing, and instrumentation. A key subproblem of MSA is section boundary detection---determining whether a given point in time marks the transition between musical sections. In this paper, we study automatic section boundary detection for symbolic music. First, we introduce a human-annotated MIDI dataset for section boundary detection, consisting of metadata from 6134 MIDI files that we manually curated from the Lakh MIDI dataset. Second, we train a deep learning model to classify the presence of section boundaries within a fixed-length musical window. 
Our data representation involves a novel encoding scheme based on synthesized overtones to encode arbitrary MIDI instrumentations into 3-channel piano rolls.
% In addition to note-level data, we devise an encoding scheme to incorporate higher-level MIDI information such as instrumentation and control change information. {\color{red} this isn't exactly correct, the encoding scheme is only for instrumentation but we also incorporate control change information...}. 
Our model achieves an $F_1$ score of $0.77$, improving over the analogous audio-based supervised learning approach and the unsupervised block-matching segmentation (CBM) audio approach by $0.22$ and $0.31$, respectively. We release our dataset, code, and models.\footnote{Dataset available at \url{https://github.com/m-malandro/SLMS}. Code and models available at \url{https://github.com/omareldeeb/midi-msa}.}
\end{abstract}

\section{Introduction}\label{sec:introduction}

Music is commonly structured over time in a hierarchical manner, ranging from short repeating phrases and motifs to longer, non-overlapping sections such as verses, choruses, or movements. The automatic analysis of this structure is known as Music Structure Analysis (MSA) and can characterize analyses spanning a wide temporal range---from brief segments lasting a few seconds, to entire sections exceeding a minute in duration. A common first step in MSA is the detection of section boundaries, which can then be used to group or label segments based on principles such as homogeneity and repetition.
In this work, we focus on detecting section boundaries in symbolic music in a non-hierarchical manner---that is, identifying the points in time where one musical section (e.g., verse, chorus, bridge, etc.) ends and another begins.

% Due to the inherently subjective nature of music structure, human annotations are sometimes inconsistent from annotator to annotator. Nevertheless, existing methods often fall short of matching the level of agreement observed across multiple human annotators. This is due in part to the limited availability of annotated datasets \cite{NietoOverview}. 
%---This is especially true for symbolic music, where segment annotations are particularly scarce.

So far, most algorithms for MSA have focused on waveform audio as opposed to symbolic data, possibly due to a current lack of human-annotated symbolic music. Exceptions include \cite{pop909structure}, which focused on phrase-level segmentation in pop piano music 
% with annotated melody and harmony tracks 
(and proposed identifying sections from the patterns of detected phrases), and \cite{midi_seg_temporal}, which focused on phrase-level segmentation in melodies. 
% The lack of work on segmentation of arbitrary symbolic music is perhaps due to the lack of annotated symbolic data, which we help alleviate in this paper.

MSA of waveform audio is a central research topic in music information retrieval, and is motivated by a number of downstream applications \cite{NietoOverview}. Here we highlight two motivations for the study of MSA for symbolic music: 

First, for researchers who are interested in MSA, symbolic data are more freely and openly available than waveform data. While existing annotations of waveform audio are generally freely and openly available \cite{SALAMI, Harmonix, rwc, rwc2, rwc3}, obtaining access to all of the associated audio recordings can be expensive. 
% Additionally, different releases of audio recordings may have different amounts of silence at the beginning of tracks, requiring manual corrections of public annotations. 
In contrast, symbolic datasets are more freely and widely available. 
We release a new dataset, consisting of human annotations of section boundaries of 6134 MIDI files from the Lakh MIDI Dataset (LMD) \cite{lmd1, lmd2}.
% , which we found in the metadata of these files and manually verified. 
All of the MIDI referenced by these annotations is available in the LMD.

Second, MSA for symbolic music has the potential to improve the quality of symbolic music generation. Previous works have identified a trend for outputs from generative symbolic models to be repetitive or meandering, rather than having clear structure that drives toward musical payoffs \cite{whatsmissing, tradformer, jigs}. Recent work \cite{SymPAC} introduced a structure-aware symbolic music generation system, which is capable of writing contrasting musical sections. We note that the authors of \cite{SymPAC} used an audio-based method \cite{Foote}, applied to waveform data, to compute section boundaries for their training data. If MSA techniques for symbolic music outperform MSA techniques for audio, then such techniques would improve the training data quality, and therefore likely also improve the output quality, of structure-aware generative symbolic music systems.

\section{Related Work}
% \subsection{Methods}
Data for music structure analysis are scarce, as manually annotating music is labor-intensive. Audio datasets for MSA include SALAMI \cite{SALAMI}, the Harmonix dataset \cite{Harmonix}, and the RWC dataset \cite{rwc, rwc2, rwc3}. 
% These datasets contain annotations for audio recordings of mostly Western popular music (although classical, jazz, and other styles are also present). 
Annotated symbolic datasets include the piano dataset Pop909 \cite{pop909} and its annotations in \cite{pop909structure}, the Essen Folksong dataset \cite{essen} (which consists of phrase-level annotations of 8473 short folk melodies), and S3, an annotated dataset of 4 symphonies totaling 16 movements \cite{S3}. We aim to develop a method capable of segmenting arbitrary MIDI files, and therefore developed our own dataset for this work---see \secref{SecDataset} for details.

In 2014, Ullrich, Schlüter, and Grill \cite{Ullrich2014BoundaryDI} express the problem of section boundary detection in audio as a binary classification task, and train a convolutional neural network (CNN) on mel-scaled spectrograms extracted from fixed-duration slices of audio to predict whether a section boundary exists at the center of a given network input. At test time, the network is applied over a sliding window of the audio and produces for each frame a boundary probability. These boundary probabilities are finally decoded to boundary positions using a simple peak-picking algorithm with a moving threshold.

In two follow-up papers \cite{2015SSLMs, 2015BoundaryDetection}, Grill and Schlüter extend this approach by incorporating additional input representations and multi-level annotations. They use Harmonic-Percussive Source Separation (HPSS) to isolate harmonic and percussive components and Self-Similarity Lag Matrices (SSLMs) to capture long-range temporal dependencies. This approach remains the state of the art in supervised section boundary detection on the SALAMI dataset \cite{SALAMI}. Subsequent papers have explored self-supervised learning approaches \cite{2019Unsupervised}, hierarchical MSA \cite{MultiMSA2022, MultiMSA2024}, and the functional labeling of segments \cite{MSALabeling2024, to-catch-a-chorus, all-in-one}.
% More recently, Transformer-based architectures have been explored for music segmentation \cite{to-catch-a-chorus, all-in-one}. In \cite{to-catch-a-chorus}, a spectral-temporal Transformer-based model is used to simultaneously predict boundary probabilities and functional classes at the frame level and improves over Grill and Schlüter \cite{2015BoundaryDetection} on the RWC-Pop \cite{rwc} and Isophonics datasets. Similarly, \cite{all-in-one} proposes a Transformer-based model for frame-level prediction of boundary and functional label, and adds additional learning targets for beats and downbeats, which results in a unified model with state-of-the-art results across all four tasks on the Harmonix dataset \cite{Harmonix}.
More recently, Transformer-based models have been proposed to jointly detect boundaries and functional labels \cite{to-catch-a-chorus, all-in-one}, achieving state-of-the-art results on datasets like Harmonix \cite{Harmonix}.

\begin{figure*}[t]
    \centering
    \includegraphics[alt={A patch (model input), showing contrast between the musical content to the left and to the right of the center of the patch.},width=.9\linewidth]{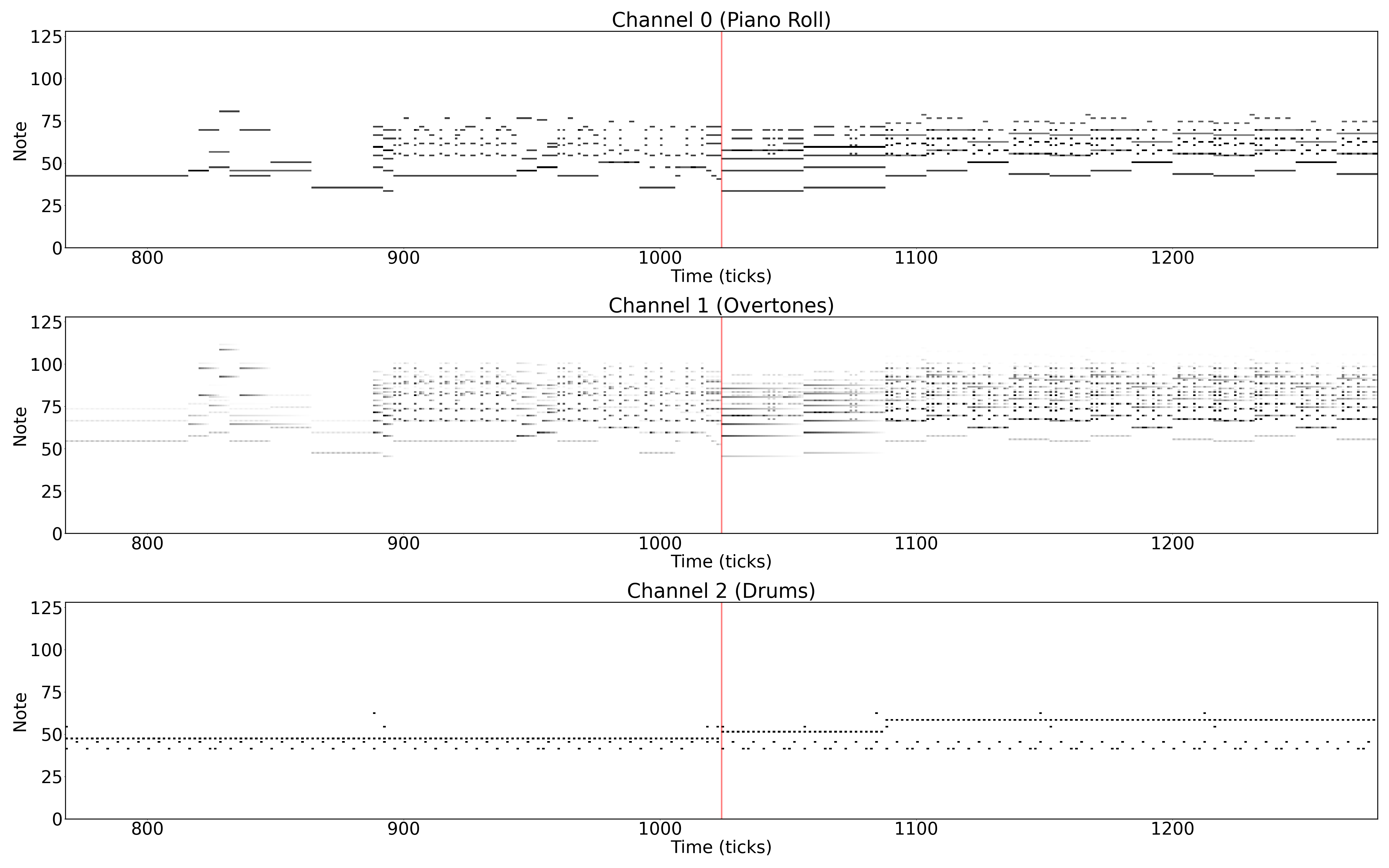}
    \caption{A patch from file ca05cc474fd2010484c1201bf57b3cfd from the training set. We overlay the vertical red line in the center of the patch in this figure to indicate that this is a positive (section boundary = True) training example.}
    \label{fig:example_input}
\end{figure*}

\section{Method}

\label{SecMethod}

Our method is inspired by the audio-based method in \cite{2015BoundaryDetection} and the line of work leading up to it \cite{Ullrich2014BoundaryDI, 2015SSLMs}. In particular, our method involves training a convolutional neural network (CNN) on piano rolls synthesized from MIDI data.

\subsection{Feature Extraction}
Given a MIDI file, we extract the time, pitch, duration, velocity, and program (instrumentation) information of each note event. Additionally, we multiply the note's velocity by any preceding volume and expression control change values (scaled to the interval $[0, 1]$) on the corresponding message channel.
% , to represent the loudness of the note during playback.
We quantize all events to a fixed temporal resolution of 4 ticks per beat and merge all tracks---except for drums---into a single "piano roll" image.
We split each piano roll into equal-size patches centered at measure boundaries, which serve as the primary inputs to our neural network. The patches have a height of 128 pixels, corresponding to the 128 MIDI pitch values, and a width of 512 pixels, corresponding to a duration of 128 quarter notes, or 32 bars in a 4/4 time signature. We emphasize that our inputs do not have to be in 4/4---we accommodate files containing any collection of time signatures, and we compute measure boundaries from the time signature information within the files.
As in \cite{2015BoundaryDetection}, we separate drums into a distinct channel in order to allow the network to easily distinguish between rhythmic and harmonic/melodic content. Furthermore, we disregard the duration given by drum note events and set them to an arbitrary but fixed duration of one 16th note.
% , providing the network with an independent, more abstract representation of the excerpt's rhythm.

\subsection{Instrument Encoding}\label{subsec:overtones}
We hypothesize that explicitly encoding instrumentation (i.e., MIDI program numbers) into the piano roll representation simplifies the learning task. Given that MIDI defines a fixed number of 128 possible programs, a naive approach would be to assign each program its own input channel. However, in most cases this results in extremely sparse tensors, as most musical pieces use only a small number of instruments. Instead, inspired by audio spectrograms, we propose a harmonic overtone encoding scheme:
\begin{itemize}
    \item Each non-drum program is mapped to a random but fixed ``harmonic overtone series.''
    \item Each played note generates $K$ additional overtones (with decreasing velocity factors) at integer multiples of the original note's fundamental frequency, up to a maximum multiple of~$5$.
     
     For example: Piano programs could be mapped to the sequence $(2, 3, 5)$ with velocity factors $(0.6, 0.4, 0.1)$. If any piano track plays a note with fundamental frequency $f_0$ and velocity $v$, three additional notes at frequencies $(2f_0, 3f_0, 5f_0)$ are generated, quantized to the closest MIDI note, and added to the piano roll with onset velocities $(0.6v, 0.4v, 0.1v)$.
    \item We apply a linear amplitude decay to each generated overtone over the note’s duration, scaling its velocity from full strength at note onset down to zero at the note’s end. This helps distinguish overtones from actual note-onset events.
    \item The overtone-based encodings are assigned to a dedicated input channel, separate from the primary piano roll representation.
\end{itemize}

\subsection{Model Architecture}\label{sec:architecture}
CNNs have achieved state-of-the-art results across various MIR tasks, including section boundary detection for audio \cite{2015BoundaryDetection}, onset detection \cite{6854953}, and beat tracking \cite{8902578}. Given their success---particularly for the closely related task of audio-based section boundary detection---we adopt a CNN-based approach for our symbolic music task.

Recent work in MIR has demonstrated the effectiveness of finetuning CNNs pretrained on computer vision tasks for music-related applications \cite{lattner_samplematch_2022, arguello_cue_2024}. Inspired by these findings, we use MobileNetV3 \cite{howard2019searching}, a lightweight yet effective CNN architecture originally designed for efficient image recognition. 
% We adapt MobileNetV3 by replacing the classification head to match the number of target outputs defined by our learning task (see \secref{sec:multi_task_learning}) and 
We train the model starting from weights pretrained on the ImageNet dataset \cite{ImageNet2009}.
We demonstrate in \secref{SecExperiments} that this approach yields slightly better results than training the same architecture from scratch.

We adopt the learning task outlined in \cite{2015SSLMs, 2015BoundaryDetection, Ullrich2014BoundaryDI}, where the network aims to predict whether a section boundary is present at the center of the input patch. An example of an input to our neural network is given in \figref{fig:example_input}.

\section{Dataset}
\label{SecDataset}

The Lakh MIDI dataset (LMD) \cite{lmd1, lmd2} is a dataset of approximately 170k MIDI files widely used by the research community.
In this section, we introduce a new subset of the LMD, which we call the Segmented Lakh MIDI Subset (SLMS). 

% We became interested in section boundary detection for symbolic music due to its potential to improve symbolic music generation. However, while we found previous work on phrase-level segmentation for symbolic music \cite{pop909structure, midi_seg_temporal}, we found no work on section boundary detection for symbolic music, possibly because---aside from the recently-released S3 dataset \cite{S3} (consisting of 4 annotated symphonies)---we also found no datasets to support such work. We therefore decided to look through existing large symbolic music datasets to see if any happened to contain section-level annotations. 

While exploring the LMD, we noticed that thousands of the files contain MIDI markers. The ``marker'' event in MIDI is a meta event with a text string field that can be placed at any time location in a MIDI file. In some of these files, the markers are intended by the MIDI file authors to be section boundary markers, and in some of these, we found these markers to serve as reasonable segmentations. 
The authors of \cite{2015BoundaryDetection} point out that there can be a wide range of opinion in how to segment a piece of music---evaluating the section-level dual annotations in SALAMI against each other with an evaluation tolerance of 0.5 seconds, they found an F1 score of 74\%. Therefore, we judge annotations to be segmentations whenever we find them to be reasonable, rather than whether they agree with how we would have segmented the file.

As far as we know, the existence of files containing annotated segmentations within the LMD has gone unnoticed by other researchers until now. Indeed, in the thesis in which the LMD was introduced, Raffel \cite{lmd1} found no files containing structural annotations, possibly because he searched for ``text'' events rather than ``marker'' events.
 
We began by deduping the LMD using the method in \cite{CA}, which uses silence removal and quantized note onset chromagrams to identify when two files contain essentially the same musical information. 
% Many files contain only a fraction of the markers needed to serve as a proper segmentation---for instance, we found many files where the author only put markers in the first half of a file. Other files contain segmentations {\em along with} other markers that are not intended to serve as segment boundary markers.
To filter files that are unlikely to include valid segmentations, we then excluded files that had fewer than 3 markers and files that had an unreasonably low (less than 6) or high (more than 24) ratio of measures to markers. We also excluded files which had no markers between the first and last note onsets. 
% Of the resulting files which contain segmentations, the vast majority contained section-level (as opposed to phrase-level) annotations. 

We then found a single MIDI file author (Benjamin Robert Tubb) whose name appeared in a majority (about 57\%) of the files with markers. Tubb sequenced primarily 19th century popular songs, and files bearing his name have a distinctive layout. We therefore present the SLMS as two non-overlapping subsets: The {\em Tubb files} and the {\em non-Tubb files}. The non-Tubb files are more diverse in terms of both annotation style and musical style, and consist of styles including, but not limited to, rock, metal, jazz, solo classical piano, symphonic, and karaoke music. We also identified 5 files without Tubb's name in them that we believe he sequenced. We include them in the Tubb files.

% Salami 1359 public master branch 3/11/2025
\begin{table}[t]
    \centering
    \begin{tabular}{|c|c|c|}
    \hline
       Dataset & \# Songs & \# Hours\\
    \hline
      SLMS (Tubb)  & {\bf 3907} & \bf{225.1} \\
      SLMS (non-Tubb)   & {\bf 2227} & \bf{143.5} \\
      \hline
      SALAMI & 1359 & 105.8 \\
      Harmonix & 912 & 56.1 \\
      RWC & 315 & 23.5 \\
    \hline      
    \end{tabular}
    \caption{Some annotated multi-track music datasets}
    \label{tab:datasetinfo}
\end{table}

This left us with 4466 Tubb files and 3336 non-Tubb files. We then performed a manual inspection of each of these files. We visualized each MIDI file in a digital audio workstation (DAW) to see if the markers appeared to represent a valid segmentation, and when we were unsure, we listened. We selected only files where all perceived segment boundaries occurred at bar lines (as defined by the time signature information within the files). 
% We note that, for common musical structures combined with section boundaries near short pickups, we consider the start of the measure following the pickup to be the section boundary. 
After removing files containing clear errors (such as missing markers or misplaced markers) or markers that are not intended to serve as segment boundary markers, we were left with 3907 Tubb files (12.5\% rejection rate) and 2227 non-Tubb files (33.2\% rejection rate). We quantized all markers to bar lines and extracted all resulting marker information from these files, both in terms of seconds and in terms of beats elapsed since the start of the file. We release this information as the SLMS, including our train/validation/test split.\footnote{https://github.com/m-malandro/SLMS} 

This release constitutes the largest collection of human-segmented multi-track music of which we are aware. See \tabref{tab:datasetinfo} for information about how our dataset compares to other datasets of human-segmented music. 
% We note that our full dataset and all information needed to replicate our experiments is freely and publicly available, while parts of these other datasets (specifically, much of the underlying audio) are not. 
We mention that the section boundary markers in many of these files also contain structural information (e.g., ``verse'', ``chorus'') in their text labels, which we have also extracted and included in the SLMS. This information may be useful for future work in Music Structure Analysis. 

Despite the larger size of our dataset, we also acknowledge some shortcomings of our dataset relative to others. For example, SALAMI was created with a style guide for annotators, and contains multi-level annotations. Our dataset has neither property. Also, a majority of the files in SALAMI have two human annotations, while each of our files has only one---the annotation of the original MIDI file author.

Aside from one programmatic correction we discuss below, while creating our dataset, we resisted the urge to correct segmentations that we found to be incorrect. Instead, we chose simply to exclude files containing incorrect segmentations from the SLMS. We wanted our data contribution in this work to be primarily a record of what is already present in the LMD, rather than our subjective corrections to that data. We note that we excluded many files due to simple errors or omissions, and this provides an opportunity in future work to obtain a large amount of additional training data at the expense of some additional data correction effort. We release our starting list of 7802 files along with our final list of 6134 curated files in case other authors wish to carry out this work. 

We made one programmatic correction to the Tubb files. Tubb often split measures at section boundaries containing pickups into two measures (e.g., a measure of 6/8 with a pickup beginning at the fifth eighth note would be split into a measure of 5/8 and a measure of 1/8, with the segment marker placed at the start of the 1/8 measure, rather than at the start of the next measure). For the Tubb files in our dataset, we moved markers forward to the start of the next measure when they occurred at a measure of less than half note duration followed by a measure with greater than or equal to a half note duration. Based on our experience looking at and listening to the files, we did not find this correction to introduce any segmentation errors. We note that we changed only the boundary marker locations, not the measure annotations themselves. There are many examples of short odd-time-signature measures near segment boundaries in our non-Tubb files, indicating that our model needs to be able to handle music with such embedded measure annotations to generalize to unseen data.

\section{Experiments}
\label{SecExperiments}
% \subsection{Evaluation}

For all experiments, we train a MobileNetV3\cite{howard2019searching} architecture using a binary cross-entropy loss function and optimize using AdamW \cite{AdamW} with a learning rate of $10^{-3}$ and weight decay of $10^{-2}$. We apply early stopping when no improvement in validation $F_1$ score is observed for $5$ consecutive epochs. 

We use 5359 songs for training---3425 of the Tubb files and 1934 of the non-Tubb files. We use 246 Tubb and 100 non-Tubb songs for validation. Hence, our test set contains 236 Tubb and 193 non-Tubb songs. To ensure consistency in both training and evaluation, we exclude boundaries that occur near the beginning or end of a piece. Specifically, we disregard all segment boundaries that fall within 16 bars of the first or last note onset. While prior work (e.g., \cite{Ullrich2014BoundaryDI, 2015BoundaryDetection, 2015SSLMs}) handles edge cases by padding the input with half a patch window, we observed inconsistencies in segment annotations near the beginnings and endings of some pieces in our dataset (e.g., whether the beginning of the final measure is marked as a segment boundary) and therefore instead adopt this boundary exclusion strategy, applying it uniformly to our method and all baselines. 
Hence, this paper focuses on identifying section boundaries in the ``middle'' of songs. Inconsistent boundary annotations near the beginnings and ends of songs in our dataset can be addressed in future work.

\subsection{Our Method}

We apply our piano roll-based method from \secref{SecMethod} to our dataset, using $K=3$ overtones per note. Since section boundary events in our dataset are relatively sparse, we include each positive example twice in each training epoch, while negative examples are included only once per epoch. 
% To address class imbalance resulting from the relative sparsity of section boundary events, we increase the probability {\color{red} it's not a probability though, because we are not undersampling negative examples} of sampling a positive example during training by a factor of two.

Since we are interested in barwise section boundary classification and have access to ground-truth measure positions via the MIDI files, we evaluate our approach using a per-measure hit rate---for each measure in each MIDI file, we create an input patch centered at that measure and ask the network to decide whether there is a section boundary there. We compute precision, recall, and $F_1$ score accordingly. 
% This contrasts with typical audio-based approaches, which operate higher temporal resolutions and rely on time-based tolerances to compute hit rates. 

For our approach, we consider a model's prediction to be positive when the predicted probability exceeds a fixed threshold, which we set to $0.5$ for all of our model variants. We experimented with more sophisticated post-processing techniques (including the peak-picking method of \cite{Ullrich2014BoundaryDI}), but found that they did not improve 
the $F_1$ score 
% performance
in our setting.

% In total, we are left with 428,883 example patches: 371,947 patches for training (30,611 positive and 341,336 negative), 24,684 patches for validation (2,024 positive and 22,660 negative), and the remaining 32,252 patches (2,584 positive and 29,668 negative) for testing.

\begin{table}[t]
    \centering
    % \resizebox{\linewidth}{!}{%
    \begin{tabular}{l|r|r r}
    \hline
    \textbf{Model} & $F_1$ & \textbf{Precision} & \textbf{Recall} \\
    \hline
    Ours & & & \\
    \quad ensemble & \textbf{.7838} & .8001 & .7682 \\
    \quad our model & .7675 & .7704 & .7647 \\
    \quad no pretraining & .7572 & .7078 & .8139 \\
    \quad no overtones & .7593 & .7140 & .8108 \\
    \quad no overtones, & .7661 & .7572 & .7753 \\
    \quad \quad no drum split & & & \\
    \hline
    Analogous audio & & & \\
    \quad per-measure & .5135 & .6728 & .4152 \\
    \quad 0.5s tolerance & .5523 & .5456 & .5590 \\
    \hline
    CBM \cite{CBM} (audio) & & & \\
    \quad per-measure & .4583 & .4288 & .4923 \\
    \quad 0.5s tolerance & .4488 & .4414 & .4564 \\
    \hline
    \end{tabular}
    % }
    \caption{Primary test results. $F_1$ computed from section boundaries aggregated across Tubb and non-Tubb files.}
    \label{tab:summary_results}
\end{table}

\subsection{Ablation}\label{subsec:ablation}
To isolate the contributions of individual components of our approach, we train four model variants in an incremental ablation setup. Our final model is initialized with pretrained weights (\secref{sec:architecture}) from MobileNetV3 \cite{howard2019searching}, and uses both overtone encoding (\secref{subsec:overtones}) and drum track separation in the input representation. For the ablations, we experiment with omitting only pretraining, only overtones, and both overtones and drum separation.
% % HERE
% \begin{itemize}
%     \item Ours, no overtones - drum split: Uses a pretrained model (\secref{sec:architecture}), but omits both overtone encoding (\secref{subsec:overtones}) and drum track separation in the input representation.
%     \item Ours, no overtones: Uses a pretrained model with drum separation, but no overtone encoding.
%     \item Ours - Pretraining: Trained from scratch (no pretrained weights), but includes overtone encoding and drum separation in the input.
%     \item Ours + Aux. Targets: Uses a pretrained model with both overtone encoding and drum separation, and includes the auxiliary training targets described in \secref{sec:multi_task_learning}.
%     \item Ours: Our final model---uses a pretrained model, overtone encoding, drum separation, and only the primary boundary detection target (i.e., without auxiliary targets).
% \end{itemize}
Additionally, we combined all four variants into a single bagged ensemble, averaging their output probabilities at inference time. 
% While this increases inference cost, it leads to a notable improvement in performance (see \tabref{tab:summary_results}).

% \begin{table}[t]
%     \centering
%     \begin{tabular}{c|c||c|c}
%     \hline
%         \textbf{Model} & \textbf{$\mathbf{F_1}$ score} & \textbf{Precision} & \textbf{Recall}\\
%         \hline
%         Ours (ensemble) & \textbf{.7971} & .8234 & .7724 \\
%          Ours & .7675 & .7704 & .7647\\
%          \hline
%          Ours, no pretraining & .7572 & .7078 & .8139\\
%          Ours, no overtones & .7593 & .7140 & .8108\\
%          Ours, no overtones, & .7661 & .7572 & .7753\\
%          no drum split & & & \\
%          \hline
%          USG\cite{Ullrich2014BoundaryDI}, 0.5s & .5523 & .5456 & .559 \\ 
%         % (audio, 0.5s) & & & \\
%          USG \cite{Ullrich2014BoundaryDI} & .5323 & .605 & .4752 \\ 
%         (audio, per-measure) & & & \\
%          \hline
%          CBM \cite{CBM} & .4488  & .4414 & .4564\\
%          (audio, 0.5s) & & & \\
%          CBM \cite{CBM} & .4583 & .4288 & .4923 \\ 
%          (audio, per-measure) & & & \\
%     \hline
%     \end{tabular}
%     \caption{Primary Test Results. Section boundaries aggregated across Tubb and non-Tubb...{\color{red} say something like they did in the USG? paper}}
%     \label{tab:summary_results}
% \end{table}

\subsection{Audio-based Approaches}\label{subsec:audio_comparison}

To compare our method with audio-based methods, we render the MIDI files in our dataset to audio using FluidSynth \cite{fluidsynth} and the Arachno soundfont \cite{arachno}. 

In addition to evaluating the following audio-based approaches on a per-measure $F_1$ basis, following the evaluation procedures of the Boundary Retrieval task in the Music Information Retrieval Evaluation eXchange (MIREX),\footnote{\url{https://www.music-ir.org/mirex/wiki/MIREX_HOME} accessed 27 Mar 2025} we also evaluate with $F_1$ scores with tolerances of $\pm0.5$ seconds and $\pm3$ seconds. 
% TODO {\color{red} maybe make the footnote a citation instead?}

\subsubsection{Supervised Audio Baseline}
\label{SecAudioBaseline}

For the first audio-based baseline, we implement an approach that is analogous to our MIDI-based approach and is similar to the approach in \cite{Ullrich2014BoundaryDI, 2015SSLMs, 2015BoundaryDetection}, replacing the piano rolls in our model inputs with synthesized audio. As in \cite{Ullrich2014BoundaryDI, 2015SSLMs, 2015BoundaryDetection}, we extract mel-scaled magnitude spectrograms from the synthesized audio using the same parameters. As in our symbolic approach, we separate harmonic and percussive content into distinct channels. Instead of applying HPSS as in \cite{2015BoundaryDetection}, we render drums separately from the other instruments in each file, giving us the cleanest possible separation.

As with our models, we train a pretrained MobileNetV3 \cite{howard2019searching} on these inputs. 
We attempted to train the model only on patches centered on measure boundaries, as we did for our model, but the training did not converge. Hence, we trained on patches centered on all time frames of the input spectrogram as in \cite{Ullrich2014BoundaryDI, 2015SSLMs, 2015BoundaryDetection}.
In \cite{Ullrich2014BoundaryDI}, the probability of sampling a positive example during training was increased by a factor of three. We found this to hurt the model's performance in our case, so we omit this aspect in our implementation. The remainder of the pipeline---including input resolution, training setup, and the proposed peak-picking method for boundary extraction---is kept identical to \cite{Ullrich2014BoundaryDI}, with the exception of model bagging, which we also omit. Therefore, a fair comparison is between this model and any of our individual models.

For evaluation on a per-measure basis, we provide the trained model only with inputs centered at measure boundaries. In this setting, we evaluated both thresholding and peak-picking, and found peak-picking to perform best on the non-Tubb files in our validation set. Hence, we use peak-picking to post-process the outputs of this model for all reported results.

\begin{table*}[t]
    \centering
    \begin{minipage}[t]{0.48\linewidth}
    \centering
    \begin{tabular}{l|r|r r}
    \hline
    \multicolumn{4}{c}{\textbf{Non-Tubb files}} \\
    \hline
    \textbf{Model} & ${F_1}$ & \textbf{Precision} & \textbf{Recall} \\
    \hline
    Ours & & & \\
    \quad ensemble & \textbf{.7160} & .7320 & .7007 \\
    \quad our model & .6981 & .7015 & .6947 \\
    \quad no pretraining & .6974 & .6413 & .7644 \\
    \quad no overtones & .6893 & .6415 & .7449 \\
    \quad no overtones, & .6905 & .6879 & .6931 \\
    \quad\quad no drum split & & & \\
    \hline
    Analogous audio & & & \\
    \quad per-measure & .4435 & .6729 & .3309 \\
    \quad 1 bar tolerance & .4635 & .7031 & .3457 \\
    \quad 0.5s tolerance & .4466 & .4440 & .4493 \\
    \quad 3s tolerance & .6274 & .6207 & .6342 \\
    \hline
    CBM \cite{CBM} (audio) & & & \\
    \quad per-measure & .5436 & .4994 & .5962 \\
    \quad 1 bar tolerance & .6525 & .6001 & .7150 \\
    \quad 0.5s tolerance & .4856 & .4634 & .5101 \\
    \quad 3s tolerance & .6290 & .6010 & .6597 \\
    \hline
    \end{tabular}
    \end{minipage}
    \hfill
    \begin{minipage}[t]{0.48\linewidth}
    \centering
    \begin{tabular}{l|r|r r}
    \hline
    \multicolumn{4}{c}{\textbf{Tubb files}} \\
    \hline
    \textbf{Model} & $F_1$ & \textbf{Precision} & \textbf{Recall} \\
    \hline
    Ours & & & \\
    \quad ensemble & \textbf{.8559} & .8722 & .8401 \\
    \quad our model & .8413 & .8434 & .8393 \\
    \quad no pretraining & .8234 & .7844 & .8665 \\
    \quad no overtones & .8358 & .7951 & .8809 \\
    \quad no overtones, & .8457 & .8288 & .8633 \\
    \quad\quad no drum split & & & \\
    \hline
    Analogous audio & & & \\
    \quad per-measure & .5678 & .6728 & .4912 \\
    \quad 1 bar tolerance & .7730 & .9159 & .6687 \\
    \quad 0.5s tolerance & .6424 & .6313 & .6538 \\
    \quad 3s tolerance & .7911 & .7712 & .8120 \\
    \hline
    CBM \cite{CBM} (audio) & & & \\
    \quad per-measure & .3718 & .3544 & .3911 \\
    \quad 1 bar tolerance & .6628 & .6321 & .6966 \\
    \quad 0.5s tolerance & .4105 & .4171 & .4041 \\
    \quad 3s tolerance & .6919 & .7040 & .6802 \\
    \hline
    \end{tabular}
    \end{minipage}
    \hfill
    \caption{Breakdown of test results between Tubb and non-Tubb files.}
    \label{tab:main_results}
\end{table*}

\subsubsection{Unsupervised Audio Baseline}
\label{SecCBM}

For our second audio-based baseline we use the correlation block-matching (CBM) segmentation algorithm \cite{CBM}, which is competitive with \cite{2015BoundaryDetection} on the RWC Pop dataset \cite{rwc} and marginally worse than \cite{2015BoundaryDetection} on SALAMI \cite{SALAMI}. 

The CBM algorithm requires two parameters: the number of bands $n$ and the penalty weight $w$ to the modulo-8 penalty function. The CBM algorithm also requires as input the list of bar onset times, which we provide as extracted from our MIDI files. We performed a grid search with $n\in \{7, 15\}$ and $w\in\{0, 0.04, 0.25, 0.375, 0.5, 0.75, 1\}$ and found that $n=15, w=0.25$ works best on the non-Tubb files in our validation set.
% (and works well on the Tubb files in our validation set as well, with a one-bar tolerance). 
We therefore apply the CBM algorithm with these parameters to the rendered audio of our test set. We also apply the CBM algorithm to our test data without providing the list of bar onset times, instead using the default bar-detection algorithm from their code (specifically, the downbeat estimator from the \texttt{madmom} toolbox \cite{madmom} together with the bar tracking model from \cite{MeasureTransitions}).

\subsection{Results and Discussion}

An overview of results is given in \tabref{tab:summary_results}, with a more detailed breakdown between the Tubb and non-Tubb files presented in \tabref{tab:main_results}. In these tables, ``Analogous audio'' refers to the supervised audio baseline described in \secref{SecAudioBaseline}, and ``CBM'' refers to the unsupervised audio baseline described in \secref{SecCBM}.
% For completeness, we report both time-tolerance-based and per-measure evaluation results in our comparison with audio-based approaches.
% (which we describe in \secref{subsec:audio_comparison}). 

Among our ablations, removing either overtone encoding or pretraining results in slight drops in performance. Interestingly, the variant without both overtone encoding and drum separation performs only marginally worse than the full model, suggesting that the core piano roll representation already provides a strong foundation. Performance on the Tubb files is higher than on the non-Tubb files for our models. As discussed in \secref{SecDataset}, the relative style homogeneity of the Tubb subset, as well as its higher representation in the training set, likely contribute to these results.
Overall, these results indicate that while each component contributes incrementally to performance, even the simpler variants of our approach outperform strong baselines. Moreover, as in \cite{Ullrich2014BoundaryDI}, ensemble averaging provides a practical and effective strategy to boost performance.

The fairest comparison between our method and the audio-based baselines is on the non-Tubb files in our test set, as these represent a wide range of musical genres and annotation styles, and therefore likely better represent generalizability to unseen data. 
% We use the standard $F_1$ metric (the harmonic mean of precision and recall) as our evaluation metric. 
As a secondary comparison, we compare results on the Tubb files in our test set as well. 
% Aggregate results on the Tubb and non-Tubb files are presented in \tabref{tab:summary_results}, while a breakdown of results is given in \tabref{tab:main_results}.

Our model outperforms both audio-based baselines on both the non-Tubb and Tubb files in our test set. On the non-Tubb files, the CBM algorithm outperforms the supervised audio approach, with $F_1$ scores of 0.5436 and 0.4435, respectively. This $F_1$ score obtained by the CBM algorithm is in-line with the results obtained by its authors in \cite{CBM}, between their results (with a 0.5 second tolerance) on the SALAMI (0.42) and RWC Pop (0.64) datasets. When not supplying bar onset times to the CBM algorithm, performance was evaluated using 0.5 second and 3 second tolerances, and is similar to the performance obtained by supplying the bar onset times. On the Tubb files, the supervised audio baseline outperforms the CBM algorithm, with $F_1$ scores of 0.5678 and 0.3718, respectively. The performance of our approach was considerably higher, with an $F_1$ score of 0.7675 on the non-Tubb files and 0.8413 on the Tubb files in our test set. 

We note that even if we apply loose tolerances to the outputs of our audio-based baselines (specifically, 1-bar or 3-second tolerances), the resulting $F_1$ scores are still below those achieved by our approach with strict tolerance.

\section{Conclusion and Future Work}

We have introduced a new symbolic music dataset (the SLMS) for Music Structure Analysis, containing 6134 human-annotated MIDI files. We extracted and manually curated this dataset from the Lakh MIDI dataset. We used this dataset to train a CNN to detect section boundaries in symbolic music, and have shown that our network outperforms both the analogous audio-based learning approach and the competitive correlation block-matching segmentation algorithm. 

% Our work was based on adapting the audio-based methods in \cite{2015BoundaryDetection, Ullrich2014BoundaryDI, Ullrich2014BoundaryDI} to MIDI data. The only differences between the method from \cite{Ullrich2014BoundaryDI} and the state-of-the-art audio-based approach in \cite{2015BoundaryDetection} are the use of two-level annotations (which are not available in our dataset) and the use of self-similarity lag matrices (SSLMs) as additional inputs to the model. In future work, we plan to clean and expand the SLMS via manual corrections,
Our work was based on adapting the audio-based methods in \cite{2015BoundaryDetection, 2015SSLMs, Ullrich2014BoundaryDI} to MIDI data. The ideas from the audio-based approach in \cite{2015BoundaryDetection} that we have not yet explored are the use of two-level annotations (which are not present in our dataset) and the use of SSLMs as additional inputs to the model \cite{2015SSLMs}. Based on our results in this work and the improvement in \cite{2015BoundaryDetection} over previous audio-based approaches, we do not expect that incorporating SSLMs into the audio-based supervised approach implemented in this paper would close the wide gap between its performance and the performance of our models. However, developing MIDI-based SSLMs may improve the performance of our models. 
In future work, we plan to clean and expand the SLMS via manual corrections,
% train larger models,
explore alternative model architectures,
and explore the use of MIDI-based SSLMs as model inputs.
\bibliography{MIDISegBib}

\end{document}